\documentclass{osa-article}
\journal{oe}

\newcommand{\br}{\mathbf{r}}

\newcommand{\pwr}{{\mathscr{P}}}
\newcommand{\snr}{{\cal{S}}}

\newcommand{\cur}{{\mathscr{J}}}
\newcommand{\noi}{{\mathscr{N}}}

\DeclareMathAlphabet\mathbfcal{OMS}{cmsy}{b}{n}

\begin{document}

\title{Sub-Rayleigh characterization of a binary source by spatially demultiplexed coherent detection}

\author{Chandan Datta,\authormark{1,*} Yink Loong Len,\authormark{1} Karol Łukanowski,\authormark{1,2}
Konrad Banaszek,\authormark{1,2} and Marcin Jarzyna\authormark{1}}

\address{\authormark{1}Centre for Quantum Optical Technologies, Centre of New Technologies,
University of Warsaw, Banacha 2c, 02-097 Warszawa, Poland}
\address{\authormark{2}Faculty of Physics, University of Warsaw, Pasteura 5, 02-093 Warszawa, Poland}
\email{\authormark{*}c.datta@cent.uw.edu.pl} %% email address is required



\begin{abstract}
We investigate theoretically coherent detection implemented simultaneously on a set of mutually orthogonal spatial modes in the image plane
as a method to characterize properties of a composite thermal source below the Rayleigh limit.
A general relation between the intensity distribution in the source plane and the covariance matrix for the complex field amplitudes measured in the image plane is derived.
An algorithm to estimate parameters of a two-dimensional symmetric binary source is devised and verified using Monte Carlo simulations
to provide super-resolving capability for high ratio of signal to detection noise (SNR).
Specifically, the separation between two point sources can be meaningfully determined down to $\textrm{SNR}^{-1/2}$ in the length unit determined by the spatial spread of the transfer function of the imaging system. The presented algorithm is shown to make a nearly optimal use of the measured data in the sub-Rayleigh region.
\end{abstract}

%%%%%%%%%%%%%%%%%%%%%%%%%%  Body  %%%%%%%%%%%%%%%%%%%%%%%%%%

%-----------------------------------------------------------------------------------------

\section{Introduction}

When composite incoherent light sources are imaged using a conventional optical system, the system transfer function introduces partial spatial coherence of the optical field in the image plane \cite{MandelWolf}. This coherence escapes detection with direct imaging, i.e.\ a spatially resolved measurement of the intensity distribution, which results in the well-known Rayleigh curse, where spatial resolution is lost below the scale defined by the characteristic spread of the transfer function \cite{Rayleigh1879}. As discussed in a number of recent works \cite{TsangPRX2016,NairPRL2016,LupoPRL2016,PaurOPT2016,YangOPT2016,YangPRA2017,ThamPRL2017,AngPRA2017,
LarsonOPT2018,LuNPJ2018,HradilOPT2019,BoucherOptica2020}, the Rayleigh limit can be overcome by measuring the optical intensity carried by one or more spatially extended modes in the image plane. The specific form of these modes is determined by the shape of the transfer function. Such a measurement effectively utilizes the spatial coherence information, as filtering out a spatial mode can be viewed as a coherent combination, with certain weights, of the field amplitudes across the image plane. However, it is recognized that superresolution techniques that are based on \emph{spatially demultiplexed intensity measurements} require prior knowledge of the source properties, e.g. its centroid, or, in the absence of thereof, suffer from trade-offs in the precision attainable when estimating multiple parameters characterizing the source \cite{ChrostowskiIJQI2017,RehacekPRA2017,Grace2020}. Attention also needs to be paid to imperfections, such as intermodal crosstalk \cite{GessnerPRL2020,Sorelli2021}, misalignment \cite{AlmeidaPRA2021,Sorelli2021} and dark counts \cite{LenIJQI2020,LupoPRA2020,OhPRL2021,Sorelli2021}.

The spatial coherence of light can be also measured by means of homodyne or heterodyne coherent detection.
Then, rather than with intensity measurements, one can consider spatial mode demultiplexing followed with coherent detection to access the coherence information, using, in the simplest scenario, a single mode\cite{YangOPT2016,YangPRA2017}, or most generally, a set of mutually orthogonal modes. For the latter, such a measurement strategy is becoming feasible owing to the developments in spatial light conversion technology \cite{LabroilleOpEx2014, Fontaine2019} combined with coherent detection techniques for high-capacity optical communications \cite{RyfCLEO2019}.

The purpose of this paper is to investigate the capability of \emph{spatially demultiplexed coherent detection} to characterize and determine the features of a composite source in the sub-Rayleigh regime.
In particular, we establish a general result for the spatial coherence information obtainable with this technique, as summarized in the covariance matrix for the complex amplitudes of the demultiplexed modes. Using this information, we then present an algorithm designed to estimate the source centroid, orientation, and the separation between the source points, for  the generic model of a two-dimensional binary source comprising two equally bright points whose separation is well below the Rayleigh criterion. Importantly, the set of demultiplexed modes can be quite arbitrary, and does not need to match the transfer function of the imaging system, as the latter enters only the estimation algorithm. Nevertheless, when some coarse knowledge of the source centroid and the transfer function is known, the number of modes that need to be detected can be reduced substantially, as most information about the source will be contained in these few modes. This makes the scheme presented here more practical compared to the previously analyzed one-dimensional model of spatially resolved homodyne detection \cite{DattaPRA2020} that would produce vast amounts of raw data in the case of two-dimensional array detectors.
%Hence the latter scenario may be restricted to the proof-of-principle domain.
Furthermore, we verify that the performance of the estimation algorithm presented here is close to optimal, namely that it nearly saturates the Cram\'{e}r--Rao bound for the parameters of interest in the sub-Rayleigh region.

This paper is organized as follows. In Sec.~\ref{Sec:Model} we introduce the model of the imaging system with spatially demultiplexed coherent detection and derive a general relation between the intensity distribution in the source plane and the covariance matrix for the complex amplitudes of the demultiplexed modes detected in the image plane. This result is specialized to the case of a binary symmetric source in Sec.~\ref{Sec:BinarySource}, where also the super-resolving capability of coherent detection % for high signal-to-noise ratio
is identified using the mathematical concept of the Fisher information. In Sec.~\ref{Sec:Estimation} the estimation algorithm is presented and applied to Monte Carlo simulated data. Finally, Sec.~\ref{Sec:Conclusions} concludes the paper.

\section{Model}
\label{Sec:Model}

\begin{figure}
\centering
\includegraphics[width=0.95\columnwidth]{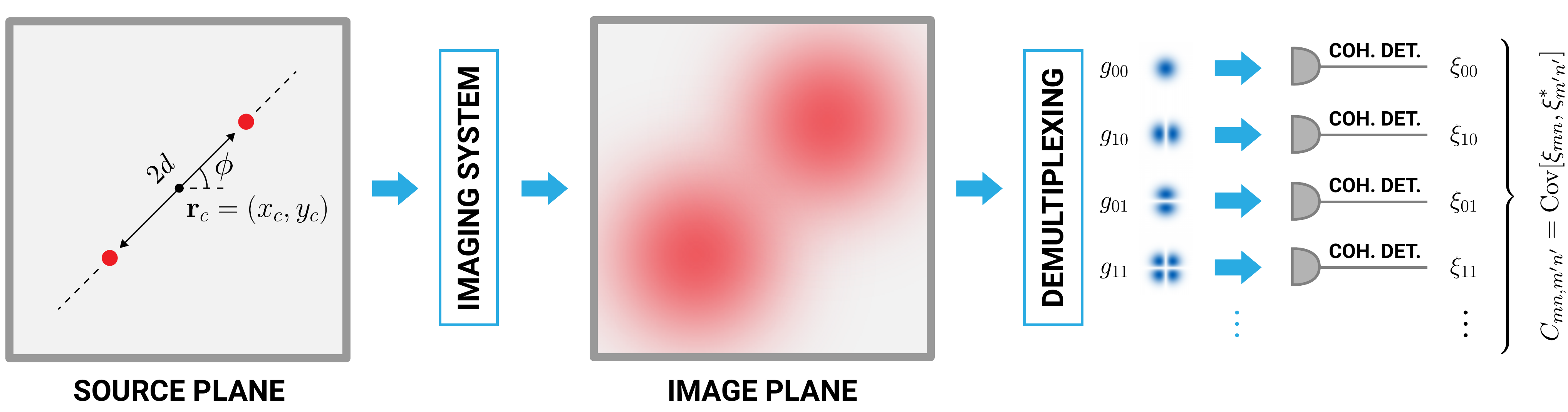}
\caption{An imaging system with spatially demultiplexed coherent detection. For concreteness, the source is taken as composed of two equally bright points separated by $2d$, oriented at an angle $\phi$, and centered at ${\mathbf r}_c = (x_c,y_c)$. The optical field in the image plane is separated into a set of orthogonal spatially extended modes $g_{mn}(\br)$ that are subject to coherent detection which yields complex amplitudes $\xi_{mn}$. The source properties are determined from the covariance matrix with elements $C_{mn,m'n'} = \textrm{Cov}[\xi_{mn},\xi^\ast_{m'n'}]$.}
\label{Fig:Model}
\end{figure}

Consider a scenario where individual points in the source plane emit mutually incoherent thermal radiation contributing to the optical field $\mathscr{E}_{\mathrm{src}}(\br)$, where the subscript $\mathrm{src}$ indicates the quantity at the source plane. Here $\br=(x,y)$ are the spatial coordinates parameterizing the plane perpendicular to the propagation axis. Under these assumptions, the coherence function of the source field can be written as
\begin{equation}
\mathbb{E}[\mathscr{E}_{\mathrm{src}}^\ast (\br) \mathscr{E}_{\mathrm{src}}(\br') ] =
\pwr w(\br) \delta(\br-\br'),
\end{equation}
where $\pwr$ %= \int d^2 \br \, \mathbb{E}[ |\mathscr{E}_{\mathrm{src}} (\br)|^2]$
is the total optical power and $w(\br)$ is the normalized intensity distribution in the source plane,
$\int d^2 \br \, w(\br) =1$.
The imaging system will be modeled by a transfer function $u(\br)$ satisfying the normalization condition $\int d^2\br \, |u(\br)|^2 =1$. Taking for simplicity unit magnification of the optical system, the signal field $\mathscr{E}(\br)$ in the image plane is expressed by
\begin{equation}
\mathscr{E}(\br)=\int d^2\br' \, \mathscr{E}_{\mathrm{src}}(\br') u(\br-\br').
\end{equation}
As depicted in Fig.~\ref{Fig:Model}, the field in the image plane is demultiplexed into a basis of orthonormal modes $g_{mn}(\br)$, labeled with a double index $mn$, that are subsequently subject to a coherent measurement of both conjugate quadratures, using e.g.\ heterodyne detection.
%%Each of the demultiplexed modes is subject to heterodyne detection as shown in Fig.~\ref{Fig:Model}.

The result of the measurement on the $mn$th mode is a complex amplitude $\xi_{mn}$ whose real and imaginary parts correspond to the two conjugate quadratures of the detected field. The detection noise will be modeled by an additive component $\cur_{mn}^{\textrm {noise}} \sim {\cal CN}(0,\noi)$ taken to follow a symmetric complex Gaussian distribution with a zero mean and a variance $\noi$ equal for all $mn$, assumed to be uncorrelated with the detected optical signal as well as between the demultiplexed modes.
For notational simplicity the amplitude $\xi_{mn}$ will be specified in units determined by the variance $\noi$ of the detection noise:
\begin{equation}
\xi_{mn} = \frac{1}{\sqrt{\noi}} \left( \int d^2 \br \, g^\ast_{mn}(\br) \mathscr{E}(\br) + \cur_{mn}^{\textrm{noise}} \right).
\end{equation}

Combining the assumption of mutually incoherent thermal radiation and others made about the source characteristics, the imaging system, and the detection scheme, it is straightforward to obtain \cite{Leonhardt1997} that the variables $\xi_{mn}$ are described by a circularly-symmetric complex multivariate Gaussian distribution with a zero mean, $\mathbb{E}[\xi_{mn}]=0$, and its covariance matrix $\mathbf{C}$ has elements equal to
\begin{equation}
\label{Eq:CovMatElements}
C_{mn,m'n'} =
\textrm{Cov}[\xi_{mn},\xi^\ast_{m'n'}]
= \mathbb{E}[ \xi_{mn} \xi_{m'n'}^\ast ] = \snr \int d^2 \br' \, w(\br')
\bar{u}_{mn}(\br') \bar{u}_{m'n'}^\ast (\br')  + \delta_{mm'}\delta_{nn'}.
\end{equation}
Here $\snr = \pwr/\noi$ is the signal-to-noise ratio and $\bar{u}_{mn} (\br')$ are projections of the transfer function displaced by $\br'$ onto the $mn$th demultiplexed mode,
\begin{equation}
\bar{u}_{mn} (\br') = \int d^2 \br \, g_{mn}^\ast (\br) u(\br-\br').
\end{equation}
One can reproduce the case of pixelated array detection by taking $g_{mn}(\br)$ equal to $1/\sqrt{\Delta x\Delta y}$ over a rectangle of dimensions $\Delta x \times \Delta y$ centered at $(m\Delta x, n\Delta y)$ and zero outside, with  $m,n$ running through the entire integer range, provided that the dimensions $\Delta x, \Delta y$ are much smaller than the spatial variation of the transfer function $u(\br)$. Such a detection scheme can be viewed as a direct, albeit noisy, measurement of the spatial coherence function in the image plane, with the noise contribution given by the second term of Eq.~(\ref{Eq:CovMatElements}).  For future reference, we also specify here a concrete example of
Hermite-Gaussian functions indexed with non-negative integers $m, n \ge 0$
\begin{equation}
\label{Eq:HermiteGaussModes}
g_{mn}(\br) = \frac{\sqrt{2}}{\sigma_D} \mathrm{HG}_m\left(\sqrt{2}x/\sigma_D\right) \mathrm{HG}_n\left(\sqrt{2}y/\sigma_D\right).
\end{equation}
Here $\sigma_D$ determines the spatial extent of the modes and $\mathrm{HG}_n(x)$ stand for the Hermite-Gaussian functions given by
\begin{equation}
\label{Eq:HGdef}
\mathrm{HG}_n(x) = \frac{1}{\sqrt{2^n n!\sqrt{\pi}}}H_n(x)\exp\left(-x^2/2\right),
\end{equation}
where $H_n(x)$ denotes the $n$th Hermite polynomial. A demultiplexing technique for mode functions specified in Eq.~(\ref{Eq:HermiteGaussModes}) has been recently demonstrated experimentally \cite{Zhou2018}.

If the set of the mode functions $g_{mn}(\br)$ is complete, knowledge of the covariance matrix elements $C_{mn,m'n'}$
makes it possible to reconstruct the variance of the coherently detected complex amplitude $\xi_v$  for any normalized mode $v(\br)$ in the image plane. The explicit expression for the variance $\mathrm{Var}[\xi_v]$ reads
\begin{equation}
\label{Eq:Varxiv}
\mathrm{Var}[\xi_v]  = \snr \int d^2\br' \, w(\br') \left| \int d^2 \br \, v(\br) u^\ast(\br-\br') \right|^2 +1 = \sum_{mn,m'n'} \bar{v}_{mn}^\ast \bar{v}_{m'n'}C_{mn,m'n'},
\end{equation}
where 
\begin{equation}
\label{Eq:vmn}
\bar{v}_{mn} = \int d^2\br \, g_{mn}^\ast (\br) v(\br).
\end{equation}
Thus, we emphasize that in principle coherent detection implemented for any choice of the set of demultiplexing modes, as long as it forms a complete basis, yields the same information about the field quadratures. Let us also note that for shot-noise limited coherent detection the signal-to-noise ratio $\snr$ is given by the mean number of photons carried by the signal \cite{LenIJQI2020,BanaszekJLT2020}.

\section{Binary source}
\label{Sec:BinarySource}

Optical coherence in the image plane introduced by the transfer function $u(\br)$ of the imaging system can provide means to identify sub-Rayleigh features of the source \cite{NairPRL2016,LupoPRL2016}. We will study this capability using the canonical example of a symmetric binary source with the normalized intensity distribution $w(\br)$ of the form
\begin{equation}
w(\br) = \frac{1}{2} [ \delta(\br - \br_1) + \delta(\br - \br_2)],
\end{equation}
where $\br_1,\,\br_2$ denote locations of light source components. It will be convenient to write $\br_1 = \br_c + d\mathbf{e}_\phi$ and $\br_2 = \br_c - d\mathbf{e}_\phi$,
where $\br_c=(x_c, y_c)$ is the source centroid, $d$ specifies the half-separation between the two source points, and the unit vector $\mathbf{e}_\phi = (\cos\phi, \sin\phi)$ determines the source angular orientation.
Henceforth we will use the soft-aperture model for the imaging system with a Gaussian transfer function given by
\begin{equation}
\label{Eq:u(r)soft}
u(\br) = \left(\frac{2}{\pi}\right)^{1/2}\exp\left(-\br^2\right)
=
\sqrt{2} \mathrm{HG}_0\left(\sqrt{2}x\right) \mathrm{HG}_0\left(\sqrt{2}y\right).
\end{equation}
The above formula assumes that the spatial spread of the transfer function defines the unit of length, meaning that centroid coordinates as well as separation appearing in equations are dimensionless quantities. Other quantities, such as parameters of the source or the spatial extent of the demultiplexed modes $\sigma_D$ used in Eq.~(\ref{Eq:HermiteGaussModes}) will be specified relative to this unit.
The second expression in Eq.~(\ref{Eq:u(r)soft}) is written using notation introduced in Eq.~(\ref{Eq:HGdef}).

The objective now is to use data collected by means of spatially demultiplexed coherent detection to estimate the centroid $\br_c$, half-separation $d$, and orientation $\phi$ of a symmetric binary source in the sub-Rayleigh regime when $d\ll1$. Motivated by the one-dimension study carried out in \cite{DattaPRA2020},
we will consider
a normalized reference mode $v_{\br_r; \phi_r}(\br)$ of the form
\begin{multline}
\label{Eq:refmodedef}
v_{\br_r; \phi_r}(\br) = \sqrt{2} \mathrm{HG}_1\left(\sqrt{2}[(x-x_r)\cos\phi_r+(y-y_r)\sin\phi_r ]\right)\\
\times \mathrm{HG}_0\left(\sqrt{2}[-(x-x_r)\sin\phi_r+(y-y_r)\cos\phi_r ]\right).
\end{multline}
The above expression can be viewed as a product of a one-dimensional transfer function $\mathrm{HG}_0$ in one direction and its normalized derivative $\mathrm{HG}_1$ in the orthogonal direction, rotated by $\phi_r$ and displaced by $\br_r=(x_r,y_r)$.
The variance $V(\br_r; \phi_r)$ of the coherently detected complex amplitude for the mode $v_{\br_r; \phi_r}(\br)$ can be calculated from the covariance matrix elements $C_{mn,m'n'}$ using Eqs.~(\ref{Eq:Varxiv}) and (\ref{Eq:vmn}) with $v_{\br_r; \phi_r}(\br)$ used in lieu of $v(\br)$ in Eq.~(\ref{Eq:vmn}).

\begin{figure}
\centering
\includegraphics[width=0.95\columnwidth]{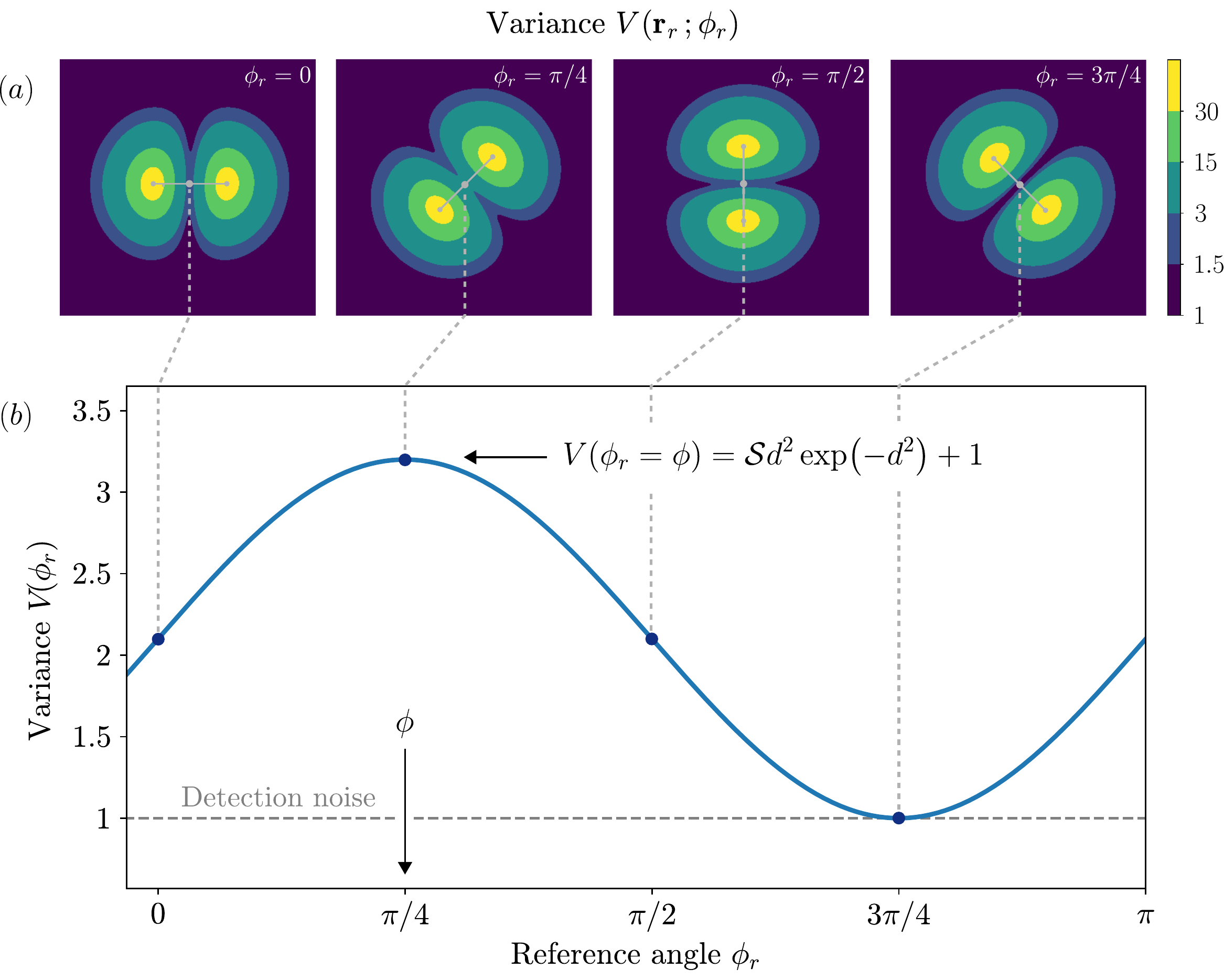}
\caption{(a) Variance $V(\br_r; \phi_r)$ of the coherently detected complex amplitude for the reference mode $v_{\br_r; \phi_r}(\br)$ defined in Eq.~(\ref{Eq:refmodedef}) as a function of the displacement $\br_r$ for several values of the rotation angle $\phi_r =0, \pi/4, \pi/2,$ and $3\pi/4$. Taking the variance value $V(\phi_r)$ at the midpoint between the maxima of the two lobes, which ideally is the centroid location, $\br_c$, yields a sinusoidal dependence on the rotation angle $\phi_r$ shown in the panel (b). The abscissal shift of the sinusoid of $V(\phi_r)$ is determined by the angular orientation $\phi$ of the binary source, whereas the amplitude of the oscillations depends on the half-separation $d$ and the signal-to-noise ratio $\snr$.
}
\label{Fig:Variance}
\end{figure}

A careful inspection of the dependence of $V(\br_r; \phi_r)$ on its arguments $\br_r$ and $\phi_r$ allows one to identify the following relation to the source parameters. As shown in Fig.~\ref{Fig:Variance}(a), for a given $\phi_r$ the graph of $V(\br_r; \phi_r)$ as a function of $\br_r$ has the form of two lobes surrounding symmetrically the source centroid. While the angular orientation of the lobes is determined by $\phi_r$,
the value of the variance at the midpoint between the lobes, i.e., $\br_c$, exhibits dependence on both $d$ and $\phi$ as depicted in Fig.~\ref{Fig:Variance}(b). The actual expression for the variance at the source centroid reads:
\begin{equation}
\label{Eq:Vbrr=brc}
V(\phi_r)=V(\br_r = \br_c; \phi_r) = \snr d^2 \exp\left(-d^2 \right) \cos^2(\phi_r-\phi) + 1.
\end{equation}
It is seen that the location of the maximum of $V(\br_r = \br_c; \phi_r)$ as a function of $\phi_r$ corresponds to the orientation of the source, whereas the source separation can be read out from the height of the maximum above the detection noise level, equal to one in the chosen units.
In the sub-Rayleigh region, when $d \ll 1$, the expression given in Eq.~(\ref{Eq:Vbrr=brc}) can be simplified
to the leading order in $d$, which yields
\begin{equation}
\label{Eq:Vbrr=brcSimplified}
V(\br_r = \br_c; \phi_r)
\approx \snr d^2 \cos^2(\phi_r-\phi) + 1.
\end{equation}

In order to gain a simple insight into how precisely the source separation can be inferred from the variance map $V(\br_r; \phi_r)$ let us recall the notion of the Fisher information. Generally, for any unbiased local estimator $\tilde{\theta}$ of a real continuous parameter $\theta$, the Fisher information ${\mathcal F}_\theta$ provides an upper bound on the attainable precision $(\Delta^2\tilde{\theta})^{-1}$, defined as the
inverse of the estimator variance $\mathrm{Var}[\tilde{\theta}]$ normalized by the sample size $N$ used for estimation:
\begin{equation}
(\Delta^2\tilde{\theta})^{-1}  = \frac{1}{N\mathrm{Var}[\tilde{\theta}]}
\le {\mathcal F}_\theta.
\end{equation}
When a parameter $\theta$ is estimated from
a univariate circularly symmetric complex normal distribution with a variance $V$, the Fisher information reads \cite{Slepian1954, Bangs1971}
\begin{equation}
\label{Eq:Fthetaunivariate}
{\cal F}_\theta = \left(\frac{1}{V} \frac{\partial V}{\partial \theta}\right)^2.
\end{equation}
For the sake of simplicity, assume for a moment that the source centroid is known perfectly and that the angle $\phi_r$ is set to $\phi$.
Using the approximate formula for $V(\br_r = \br_c; \phi_r=\phi)$ given in Eq.~(\ref{Eq:Vbrr=brcSimplified})
it is elementary to derive the simplified expression for the Fisher information:
\begin{equation}
\label{Eq:SimpleFisher}
{\cal F}_d \approx  \snr \frac{4\snr d^2}{(1+ \snr d^2)^2}.
\end{equation}
Importantly, ${\cal F}_d$ attains the value of the order of $\snr$ for $d \sim \snr^{-1/2}$. This contrasts with direct imaging, where the corresponding precision level is reached only for $d \sim 1$ in the units determined by the spread of the transfer function \cite{LenIJQI2020}. The above observation indicates the super-resolving capability of coherent detection for high signal-to-noise ratio, when $\snr \gg 1$. Note, however, that  ${\cal F}_d$ still vanishes for $d=0$, which is due to the presence of shot noise characteristic for quadrature detection \cite{LenIJQI2020, DattaPRA2020}.

\section{Estimation algorithm}
\label{Sec:Estimation}

In order to estimate simultaneously the full set of parameters of a two-dimensional binary symmetric source  the following numerical algorithm is proposed. First, within the interval $0 \le \phi_r \le \pi$, select $K$ equidistant sampling points $\phi_k = k \pi/K$, where $k=1,\ldots, K$. For each $k$, find locations of the two maxima of the variance $V(\br_r;\phi_r = \phi_k)$ as a function of $\br_r$ and take the midpoint $\br_k = (x_k,y_k)$ between these maxima. Fitting a function given on the right hand side of Eq.~(\ref{Eq:Vbrr=brc}) to the set of points $\bigl( \phi_k, V(\br_r=\br_k;\phi_r = \phi_k)\bigr)$ yields estimates for the source half-separation $\tilde{d}$ and orientation $\tilde{\phi}$. % provided that the signal-to-noise ratio value $\snr$ is known.
The estimate for the source centroid is given as an arithmetic average of all the midpoints,
$\tilde{\br}_c = (\tilde{x}_c, \tilde{y}_c) = (1/K) \sum_k \br_k $.

Importantly, note that in a practical scenario the variance $V(\br_r;\phi_r)$ is calculated from a finite set of detected modes $g_{mn}(\br)$. Because of this limitation one is able to access spatial coherence information only in a certain area of the image plane. Therefore one should possess at least a coarse knowledge of the source centroid and the transfer function in order to choose a set of demultiplexed modes $g_{mn}(\br)$ that covers the image. The better these parameters are known the lower number of modes needs to be detected.

The above estimation algorithm has been applied to a Monte Carlo simulated data for a binary source characterized by $\br_c = (-0.05, 0.1)$ and oriented at $\phi=\pi/4$. We assumed heterodyne detection implemented on a set of Hermite-Gauss modes defined in Eq.~(\ref{Eq:HermiteGaussModes}) with $\sigma_D =0.8$ and the indices $m,n$ restricted to the range $m,n \le 4$. The latter condition has been chosen to ensure that the detected modes carry nearly entire signal power. This requirement can be verified by calculating the sum
$
\sum_{m,n \le 4} \int d^2 \br' \, w(\br') |u_{mn}(\br')|^2
$
that exceeds $0.9994$ for all values of the half-separation $d$ used in the Monte Carlo simulations.
One realization of the experiment was taken to comprise $N=500$ elementary measurements of field quadratures for the chosen set of Hermite-Gauss modes. The statistical properties of the estimators were analyzed using 1000 of such realizations. The estimation quality can be further improved by taking into consideration more modes in the numerical postprocessing, which, in view of their extremely low signal amplitude, can be essentially approximated by just carrying only detection noise. Note that in the case of larger separations these higher order modes carry more information, however, in such instances one is outside the sub-Rayleigh regime and direct imaging techniques can be used instead. Specifically, the results presented below have been obtained for a heterodyne experiment simulated with Hermite-Gauss modes up to $m,n \le 4$ and then calculating the variance $V(\mathbf{r}_r;\phi_r)$ including modes $0\le m,n \le 9$, where for $4 < m \le 9$ or $4 < n \le 9$ complex amplitudes $\xi_{mn} \sim {\cal CN}(0,1)$ have been taken in the estimation algorithm. Furthermore, the interval $0 \le \phi_r \le \pi$ has been sampled with $K=10$ equidistant points.

\begin{figure}
\centering
\includegraphics[width=0.95\columnwidth]{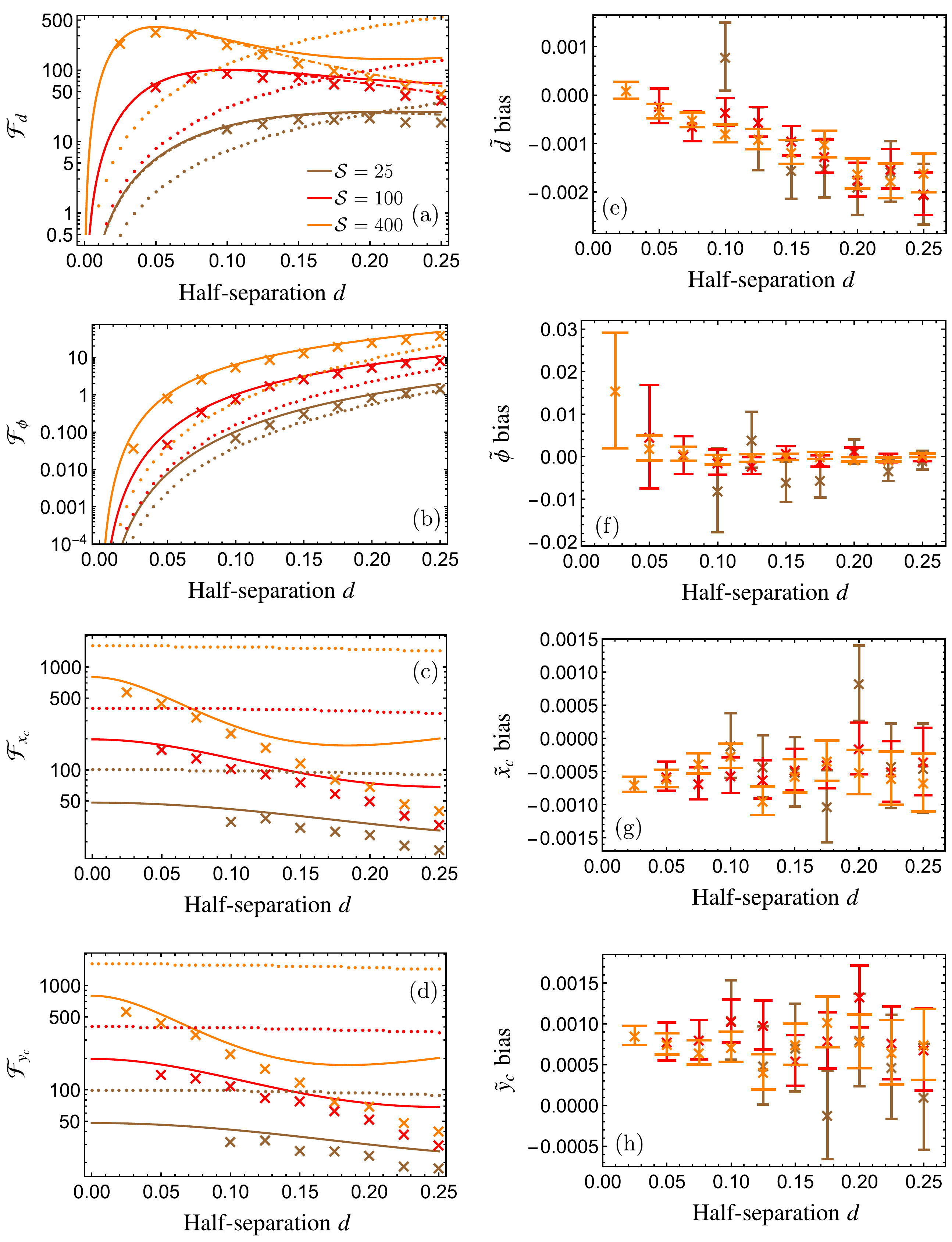}
\caption{Precision (a-d) and bias (e-h) of estimating source half-separation $d$ (a,e), orientation $\phi$ (b,f) and Cartesian coordinates of the centroid $x_c$ (c,g) and $y_c$ (d,h) for three values of the signal-to-noise ratio $\snr=25$ (brown), $100$ (red), $400$ (orange). Results of Monte Carlo simulations are shown with crosses. The solid lines in panels (a-d) depict the Cram\'{e}r-Rao bound on attainable precision obtained using Eq.~(\ref{Eq:FullCR}) whereas the dotted lines indicate Cram\'{e}r-Rao bound obtained for direct imaging. The dashed-dotted line in panel (a) represents the simplified expression derived in Eq.~(\ref{Eq:SimpleFisher}). In panels (a-d) the vertical error bars are smaller than the size of the graphic symbols representing the estimated values.}
\label{Fig:MonteCarlo}
\end{figure}

Fig.~\ref{Fig:MonteCarlo} depicts precision and bias for estimates of the source parameters: half-separation $\tilde{d}$, orientation $\tilde{\phi}$, and centroid coordinates $(\tilde{x}_c,\tilde{y}_c)$ for three values of the signal-to-noise ratio $\snr=25,100,400$ and half-separations from the range $0.025 \le d \le 0.25$ in units defined by spread of the transfer function. The dashed-dotted line in Fig.~\ref{Fig:MonteCarlo}(a) represents the simplified form of the Fisher information ${\cal F}_d$  derived in Eq.~(\ref{Eq:SimpleFisher}). It is seen that this elementary expression characterizes rather well the precision achieved by the devised algorithm for estimation of half-separation $d$. The solid lines in Fig.~\ref{Fig:MonteCarlo}(a-d) depict the Cram\'{e}r-Rao bound on attainable precision that takes into account information available in the entire covariance matrix  $\mathbf{C}$. This bound is given by the Fisher information which for data distributed according to a circularly-symmetric complex normal distribution parameterized by a parameter $\theta$ entering through the covariance matrix  $\mathbf{C}$ is equal to \cite{Bangs1971}
\begin{equation}
\label{Eq:FullCR}
{\cal F}^{\textrm{full}}_{\theta} = \textrm{Tr} \left( \mathbf{C}^{-1} \frac{\partial \mathbf{C}}{\partial\theta} \mathbf{C}^{-1} \frac{\partial \mathbf{C}}{\partial\theta} \right).
\end{equation}
It can be seen in Fig.~\ref{Fig:MonteCarlo}(a,b) that coherent detection in the sub-Rayleigh regime outperforms precision offered by direct imaging \cite{TsangPRX2016}, the latter indicated by dotted curves. Importantly, precision for orientation estimation offered by the latter detection scheme also exhibits Rayleigh limit. Unlike in the one dimensional case \cite{Grace2020} this fact therefore prevents one from using various hybrid strategies based on first estimating centroid and orientation using direct imaging and then performing estimation of separation with the appropriately adjusted SPADE measurement \cite{Grace2020}.

Note that the above expression can be viewed as a multivariate generalization of Eq.~(\ref{Eq:Fthetaunivariate}).
It is seen that the proposed estimation strategy is nearly optimal in the sub-Rayleigh region $d\ll 1$.
The loss of precision for larger value of $d$ compared to the Cram\'{e}r-Rao bound can be attributed to the need of analyzing statistics of more than just one spatial mode in the image plane \cite{DattaPRA2020}.
It can be noticed in Fig.~\ref{Fig:MonteCarlo}(e-h) that the estimates for the half-separation and the centroid coordinates exhibit a minor bias, which however remains much below the absolute values of the respective quantities and can be further reduced by including a larger number of modes.

\section{Conclusions}
\label{Sec:Conclusions}

We have investigated theoretically the capability of spatially demultiplexed coherent detection in the image plane to characterize a two-dimensional binary source composed of two mutually incoherent points. It has been found that the source separation can be estimated for values down to $\sim \snr^{-1/2}$ in units determined by spread of the transfer function of the imaging system. Here $\snr$ is the signal-to-noise ratio of coherent detection, which for shot-noise limited scenario is given by the mean number of photons reaching the image plane. This provides a super-resolving capability when $\snr \gg 1$.

We presented and tested with Monte Carlo simulations an algorithm to estimate the source parameters: its half-separation, orientation and centroid location with precision approaching the Cram\'{e}r-Rao bound in the sub-Rayleigh region. Central to the algorithm is the extraction of information about the source parameters from the variances of the quadrature measurements of the complex amplitude carried by a reference mode, for different orientation and displacement relative to the source parameters. Crucially, the choice of the reference mode can be decided later in the algorithm, and we need not to perform the experiments with the demultiplexed modes matching the imaging system transfer function---which in principle need not be known precisely at the time of the measurement---as variance in any mode can be obtained from covariance of another set of modes. Still, when coarse knowledge about the centroid and transfer function spread is given \emph{a priori}, such that the set of demultiplexed modes covers the image, the estimation algorithm performs well with just a limited number of modes, with the reference mode to use identified accordingly. This  makes spatial demultiplexing quadrature measurement more efficient than spatially resolved coherent detection which would require recording quadratures for a large number of pixels. Importantly, preliminary calculations indicate that the proposed estimation algorithm is capable of estimating source parameters with superresolving precision not only for two equally bright sources but also in more complicated scenarios involving unequal brightness.

A potentially useful aspect of spatially demultiplexed coherent detection is that in principle it yields  equivalent information about the spatial coherence even if it is implemented in a plane other than the focal plane of the imaging system. Hence an out-of-focus coherent measurement should also have a super-resolving capability for high signal-to-noise ratio. In this scenario, however, one needs to consider a complex transfer function describing the imaging system. Another interesting direction of further research is characterization of sources extended along the axis of the imaging system \cite{Zhou2019}.

\section*{Acknowledgments}
We acknowledge insightful discussions with
M. Gessner, J. Ko{\l}ody\'{n}ski, A. Lvovsky, M. Parniak-Niedojad{\l}o, and N. Treps.
This work is a part of the project ``Quantum Optical Technologies''
carried out within the International Research Agendas programme of the
Foundation for Polish
Science co-financed by the European Union under the European Regional
Development Fund. It was also supported by the US Department of Navy award no.\ N62909-19-1-2127 issued by the Office of Naval Research.

\section*{Disclosures}
The authors declare no conflicts of interest.

\section*{Data availability} 
Data underlying the results presented in this paper are not publicly available at this time but may be obtained from the authors upon reasonable request.

\end{document}